\documentclass[11pt]{article}

\usepackage{latexsym,amsmath,epsfig}

\DeclareFontFamily{OT1}{rsfs10}{}
\DeclareFontShape{OT1}{rsfs10}{m}{n}{ <-> rsfs10 }{}
\DeclareMathAlphabet{\mathscript}{OT1}{rsfs10}{m}{n}

\numberwithin{equation}{section}

\newcommand{\be}{\begin{equation}}
\newcommand{\ee}{\end{equation}}
\newcommand{\nn}{\nonumber}
\newcommand{\bea}{\begin{eqnarray}}
\newcommand{\eea}{\end{eqnarray}}

\newcommand{\ns}{\normalsize}
\newcommand{\pt}{\partial}

\def\a{\alpha}
\def\b{\beta}
\def\g{\gamma}

\def\e{\epsilon}

\def\f{\phi}

\def\k{\kappa}

\def\m{\mu}
\def\n{\nu}

\def\p{\pi}

\def\r{\rho}

\begin{document}


\begin{titlepage}

\vspace{-3cm}

\title{
   \hfill{\ns OUTP-00-06P, UPR-880T, HUB-EP-00/10\\}
   \hfill{\ns hep-th/0003256} \\[3em]
   {\huge Heterotic M--Theory Cosmology in Four and Five Dimensions}\\[1em]}
\author{
{\ns\large Matthias Br\"andle$^1$, Andr\'e Lukas$^2$ and
           Burt A.~Ovrut$^2$} \\[0.8em]
   {\it\ns $^1$Institut f\"ur Physik, Humboldt Universit\"at}\\[-0.2em]
   {\ns Invalidenstra\ss{}e 110, 10115 Berlin, Germany}\\[0.2em]
   {\it\ns $^2$Department of Physics, Theoretical Physics, 
     University of Oxford} \\[-0.2em]
      {\it\ns 1 Keble Road, Oxford OX1 3NP, United Kingdom} \\[0.2em]   
   {\it\ns $^3$Department of Physics, University of Pennsylvania} \\[-0.2em]
      {\it\ns Philadelphia, PA 19104--6396, USA}}
\date{}

\maketitle

\begin{abstract}
We study rolling radii solutions in the context of the four-- and
five--dimensional effective actions of heterotic M--theory. For the
standard four--dimensional solutions with varying dilaton
and T--modulus, we find approximate five--dimensional counterparts.
These are new, generically non--separating solutions corresponding to a
pair of five--dimensional domain walls evolving in time. Loop corrections
in the four--dimensional theory are described by certain excitations
of fields in the fifth dimension. We point out that the two exact
separable solutions previously discovered are precisely the special
cases for which the loop corrections are time--independent. Generically,
loop corrections vary with time. Moreover, for a subset of solutions
they increase in time, evolving into complicated, non-separating
solutions. In this paper we compute these solutions to leading,
non-trivial order. Using the equations for the induced brane metric,
we present a general argument showing that the accelerating backgrounds
of this type cannot evolve smoothly into decelerating backgrounds.
\end{abstract}

\thispagestyle{empty}

\end{titlepage}


\section{Introduction}

Among the simplest cosmological solutions of string theory are the
so called rolling radii solutions~\cite{mueller}. They are characterized
by a kinetic--energy driven evolution of the universe. These
fundamental solutions of string cosmology provide superinflating
cosmological backgrounds as well as subluminally expanding backgrounds
of Friedmann--Robertson--Walker type~\cite{pre_BB,ven,edrev}. 
\vspace{0.2cm}

In the present paper, we will analyze and discuss rolling
radii solutions from the perspective of the string theory/M--theory
relation. In particular, we will discuss how a superinflating phase in
string cosmology is embedded into an M--theory context. This will
be done within the framework of the four--dimensional $N=1$ effective
action of the $E_8\times E_8$ heterotic string theory and the
underlying five--dimensional effective action of heterotic
M--theory~\cite{losw1,elpp,losw2}, obtained from 11--dimensional
Ho\v rava--Witten theory~\cite{hw1,hw2,w,hor} by reduction on a
Calabi--Yau three--fold with non--vanishing G--flux.
This five--dimensional action constitutes an M--theory realization of a
``brane--world''. In precisely this context, we will analyze
rolling radii solutions and the role the
fifth dimension plays in the cosmological evolution they describe.
In particular, we will present a new class of non--separating
five--dimensional solutions that represents the direct generalization
of the well--known four--dimensional rolling radii solutions to
heterotic M--theory.

\vspace{0.2cm}

Cosmological rolling radii solutions of M--theory related to branes
have first been obtained in ref.~\cite{letter,kal0,lu,paper}. The first
cosmological solutions of five--dimensional heterotic M--theory have
been found in ref.~\cite{cosm}. These latter solutions are generalized
rolling radii solutions with an inhomogeneous fifth
dimension. Subsequently, further examples of cosmological solutions to
five--dimensional heterotic M--theory have been
presented~\cite{real}--\cite{ell}. One purpose of this paper is to clarify
the role of the solutions given in ref.~\cite{cosm} in the
present context. Potential--driven inflation
and its relation to five--dimensional heterotic M--theory has been
first analyzed in ref.~\cite{infl}. The present
paper is somewhat complementary to this work in that it addresses
similar questions, however for the case of kinetic--energy driven
inflation. Recently, there is also considerable activity, see for
example~\cite{cr}--\cite{kkop1}, exploring other
cosmological aspects of five--dimensional brane--world theories.
M--theory rolling radii cosmology based on vacua with a large number of
supersymmetries has been investigated in ref.~\cite{bfm}. In the present
paper, we consider a related situation but focus on vacua with
the ``phenomenological'' value of $N=1$ supersymmetry in four
dimensions. While this situation is of course physically favorable, we
have much less control over quantum effects than in the cases analyzed in
ref.~\cite{bfm}. In this paper, we focus on the effect of string loop
corrections in four dimensions while, for simplicity, we work at
lowest order in $\a '$. Also, we will not attempt to include
non--perturbative effects such as brane instantons. Clearly, it would
be interesting to extend the analysis presented in this paper to
include some of these effects.

\vspace{0.2cm}

Let us outline the paper and summarize its main results. To set the
stage, we first review the
relation between the four-- and five--dimensional effective actions of
heterotic M--theory as presented in ref.~\cite{losw1,infl}. In particular, we
show how the relevant four--dimensional fields, that is, the
four--dimensional metric $g_4$, the dilaton $S_R$ and the T--modulus
$T_R$ arise as moduli of the five--dimensional three--brane vacuum
solution. We also review the correspondence between excitations of the
bulk fields in the fifth dimension and string loop corrections to the
four--dimensional effective action. Correspondingly, the strong coupling
expansion parameter $\e\sim T_R/S_R$ can be interpreted as measuring 
the strength of those bulk excitations as well as the size of the loop
corrections. Then, we start with the standard class of four--dimensional
rolling radii solutions where we allow the scale factor of the
three--dimensional universe, the dilaton and the T--modulus to vary in
time. Discarding trivial integration constants, those solutions form a
one--parameter set. We then show, using the correspondence between the
four-- and five--dimensional effective theories, how this complete set
can be ``lifted up'' to approximate solutions of the
five--dimensional effective action. Due to the potentials present in
the five--dimensional theory, these solutions depend on the fifth
coordinate as well as on time and are generically non--separating.
They constitute new, non--trivial solution of the five--dimensional
effective action of heterotic M--theory that generalize the familiar
four--dimensional rolling radii solutions. More specifically,
they correspond to a pair of domain wall three--branes with
rolling radii. In addition to the overall scaling that is familiar
from four--dimensional rolling radii solutions, there is another
non--trivial feature of those solutions not visible from a
four--dimensional viewpoint.
The size of the domain wall bulk excitations (and hence the parameter
$\e$) is generically varying in time. It is this time variation of
the internal domain wall structure that makes the solutions
non--separating and, hence, non--trivial.

We can classify the
solutions according to the time--behavior of $\e$. It turns out that
there are exactly two solutions (out of the one--parameter set) for
which $\e = $ const. In those two cases, one can find exact separable solutions
which are precisely the ones that have been given in ref.~\cite{cosm}.
For all other cases, $\e$ varies in time and the corresponding exact
solution must be non--separating. As a result, the
separable solutions are exactly the ones for which loop corrections
(or equivalently five--dimensional bulk excitations) are independent
of time and are, hence, under control at all stages of the evolution.
The remaining solutions with non--constant $\e$ split into two
(one--parameter) subsets, one with increasing $\e$ and the other
with decreasing $\e$ in the negative--time branch. Particularly, the
former case of increasing $\e$ is interesting. In this case, an effectively
four--dimensional solution is subject to increasing loop corrections
that can be described by bulk excitations in the five--dimensional
theory. When $\e$ is of order one, the approximate five--dimensional
solution is no longer valid and the subsequent evolution is
described by a more complicated non--separating background. In particular,
then, the time evolution and the dependence of the fields on the additional
dimension are entangled in a complicated way. An interesting question
is whether this might help to avoid the curvature singularity at the
end of the negative--time branch. Unfortunately, no
exact analytic solution is known to us in this non--separating case.
However, we present an argument, based on the evolution equations for
the induced fields on the boundaries, that a branch change does not
occur, even at large values of $\e$. 

\section{Heterotic M-theory in four and five dimensions}

A popular starting point for string cosmology is the lowest--order
four--dimensional effective action
\begin{equation}
 S_4 = -\frac{1}{16\p
 G_N}\int_{M_4}\sqrt{-g_4}e^{-\f_4}\left[ R_4-\pt_\m\f_4\pt^\m\f_4 +\frac{3}{2}
 \pt_\m\b_4\pt^\m\b_4 +\mbox{(matter field terms)}\right]\label{S4}
\end{equation}
written in the string frame. This action can be viewed as a universal
effective action for $N=1$ compactifications (on Calabi--Yau
three--folds) of weakly coupled $E_8\times E_8$
heterotic string theory. In fact, the field content has been truncated
to the fields essential for a discussion of string cosmology, that is,
gravity and the two universal moduli $\f_4$ and $\b_4$. In terms of the
dilaton $S$ and the conventional T--modulus $T$, we can express those
fields as
\begin{equation}
 S_R = e^{\f_4}\; ,\qquad T_R=e^{\b_4}\; .\label{ST}
\end{equation}
Here $S_R$ and $T_R$ denote the real parts of the bosonic component in
the respective $N=1$ superfields. 

\vspace{0.2cm}

Given the origin of the above action, it should be possible to relate
it to the strong coupling limit of the $E_8\times E_8$ string
theory~\cite{hw1,hw2,w,hor} in its effective formulation via heterotic
M--theory, that is, M--theory on the orbifold $S^1/Z_2$. In fact, the
simplest and perhaps conceptually most interesting connection is the one
to five--dimensional heterotic M--theory~\cite{losw1,losw2}. Let us,
therefore, discuss the simplest version of this five--dimensional
theory briefly. For more details, we refer the reader to
ref.~\cite{losw1,losw2,infl}. 

This theory is obtained from its 11--dimensional counterpart by a
reduction on a Calabi--Yau three--fold with a non--vanishing G--flux.
Then the five--dimensional
space--time has the structure $M_5=S^1/Z_2\times M_4$ where $M_4$ is
a smooth $3+1$ dimensional space--time. We will use coordinates $x^\a$
with indices $\a ,\b ,\g ,\cdots = 0,1,2,3,5$ for the full five--dimensional
space--time and coordinates $x^\m$ with $\m,\n ,\r ,\cdots = 0,1,2,3$
for $M_4$. Furthermore, the $S^1$ coordinate $y\equiv x^5$ is
restricted to the range $y\in [-\p\r ,\p\r ]$ where $\r$ is the
radius of the orbicircle. In these coordinates, the action of the
$Z_2$ symmetry on $S^1$ is defined as $y\rightarrow -y$. This leads to two
four--dimensional fixed planes $M_4^1$ and $M_4^2$ at $y=0$ and
$y=\p\r$, respectively. The theory on this space--time
constitutes a five--dimensional $N=1$ gauged supergravity theory
in the bulk coupled to two four--dimensional $N=1$ theories on $M_4^1$
and $M_4^2$. A simple version~\cite{infl} of this theory is given by
\begin{multline}
 S_5 =
 -\frac{1}{2\k_5^2}\left\{\int_{M_5}\sqrt{-g}\left[R+\frac{1}{2}
 \pt_\a\f\pt^\a\f+\frac{1}{3}v^2e^{-2\f}\right]\right.\\
 +\left.\sum_{n=1}^2\int_{M_4^n}\sqrt{-g}\left[\mp 2\sqrt{2}ve^{-\f}+
 \mbox{(matter field terms)}\right]\right\}\; .\label{S5}
\end{multline}
Here $\k_5$ is the five--dimensional Newton constant, $v$ is a
constant that depends on internal instanton numbers and $\f$ is the
five--dimensional dilaton field. Its geometrical interpretation is to
measure the size of the internal Calabi--Yau space such that its
volume is proportional to $\exp (\f)$. In the same spirit as for the
four--dimensional action above, we have confined ourselves to the
field content that is essential for our cosmological discussion. The
above action constitutes an explicit realization of a
five--dimensional brane--world in the context of M--theory, as was first
realized in ref.~\cite{losw1}.

\vspace{0.2cm}

How precisely are the effective actions~\eqref{S4} and \eqref{S5}
related? This has been worked out in ref.~\cite{losw1,losw2} and we
would like to briefly review some of the results. Note that for $v\neq
0$, the action~\eqref{S5} does not admit flat five--dimensional
space--time as a solution. Instead, its ``vacuum'' is a
pair of domain walls or three--branes specified by the exact
solution~\cite{losw1}
\begin{equation}
 ds_5^2 = a_0^2Hdx^\m dx^\n\eta_{\m\n}+b_0^2H^4dy^2\; ,\qquad e^{\f} = b_0H^3
 \label{dw}
\end{equation}
with harmonic function $H=H(y)$ given by
\begin{equation}
 H=c_0+\frac{1}{3}\epsilon_0h(y)\; ,\qquad h(y) = \frac{|y|}{\p\r}-\frac{1}{2}
 \label{H}
\end{equation}
where $\epsilon_0=\sqrt{2}\p\r v$ and $a_0$, $b_0$ and $c_0$ are
arbitrary constants. This solution preserves $3+1$--dimensional
Poincar\'e invariance and represents a BPS solution of the
five--dimensional supergravity theory described by action~\eqref{S5}.
Hence, a reduction of the five--dimensional theory on this
three--brane solution to four dimensions leads to a generally
covariant $N=1$ supersymmetric theory.
This theory is, of course, the four--dimensional effective action of
the $E_8\times E_8$ string whose universal part has been given in
eq.~\eqref{S4} above. Phrased in a different way, the four--dimensional
theory provides an effective description for the moduli of the
domain wall solution. To make this explicit, define constants $\b_4$
and $\f_4$ by
\begin{equation}
 b_0 = e^{3\b_4-2\f_4}\; ,\qquad c_0=e^{\f_4-\b_4}\label{bc}
\end{equation}
as well as a four--dimensional metric $g_{4\m\n}$ by
\begin{equation}
 g_{4\m\n}=a_0^2e^{2\f_4}\eta_{\m\n}\; .
\end{equation}
Note that we can perform a general linear transformation on the
coordinates $x^\m$ in the solution~\eqref{dw}. This converts
$\eta_{\m\n}$ and, hence, $g_{4\m\n}$ into an arbitrary
four--dimensional metric. It follows that, to leading order
in $\epsilon_0$ the solution takes the form
\begin{equation}
 ds_5^2 = \left( 1+\frac{1}{3}\e h\right) e^{-\b_4-\f_4}dx^\m dx^\n
          g_{4\m\n}+\left(1+\frac{4}{3}\e h\right) e^{2\b_4}dy^2\; ,\qquad
 \f = \f_4+\e h \label{redans}
\end{equation}
where
\begin{equation}
 \e = \e_0e^{\b_4 -\f_4}\; .
\end{equation}
and $h=h(y)$ is as defined above. The metric $g_{4\m\n}$ can be
interpreted as the four--dimensional string--frame metric. The moduli
$\f_4$ and $\b_4$ measure the internal Calabi--Yau volume $V$
and the orbifold size $R$, averaged over the orbifold
coordinate~\footnote{Note in this context that the function $h(y)$
in eq.~\eqref{H} has been defined so that its orbifold average
vanishes.}. The metric and these moduli can now be promoted to
four--dimensional fields depending on $x^\m$. As discussed in
ref.~\cite{losw1,losw2},
the low--energy dynamics of these fields can be obtained by reducing
the five--dimensional action~\eqref{S5} using the
ansatz~\eqref{redans}. The resulting dynamics is precisely described
by action~\eqref{S4}. Note that this action depends on all three
moduli $a_0$, $b_0$ and $c_0$ of the exact three--brane
solution. Modulus $a_0$ is related to the scale factor in the
four--dimensional metric $g_{4\m\n}$, whereas $b_0$ and $c_0$ enter
the effective four--dimensional action through
\begin{equation}
 S_R=b_0c_0^3\; ,\qquad T_R=b_0c_0^2\; ,
\end{equation}
where we have used~\eqref{ST} and~\eqref{bc}. This establishes a
direct relationship between five--dimensional solutions
based on the three--brane~\eqref{redans} and solutions of the
four--dimensional effective action~\eqref{S4} for the moduli. Hence, via
eq.~\eqref{redans}, any cosmological solution of the four--dimensional
theory immediately implies a (approximate) cosmological solution in
five--dimensions, and vice versa.

As is apparent from the four--dimensional action~\eqref{S4}, we are
working to lowest order in $\a '$. Correspondingly, we have neglected
higher derivative terms in the five--dimensional action~\eqref{S5} as well. 
For example, to lowest order we expect $R^2$ and $R^4$ terms in the
bulk originating from the $R^4$ term in M--theory~\cite{gg} as well as
$R^2$ terms on the boundary~\cite{low}. Although it would be
interesting to include those corrections, particularly from a
five--dimensional viewpoint, for the purpose of this paper we will
focus on situations where higher--derivative corrections are still
small. 

There is another requirement for the four--dimensional effective
description to be valid. We have used a linearized approximation in
\begin{equation}
 \e = \e_0e^{\b_4 -\f_4}\sim\frac{R}{V}\sim\frac{T_R}{S_R}\label{epsilon}
\end{equation}
and, hence, $\e$ should be smaller than one for the 
action~\eqref{S4} to be sensible. What is the meaning of this
last condition? Eq.~\eqref{epsilon} leads us to three different
interpretations of the so--called strong coupling expansion parameter
$\e$. First, $\e$ measures the
excitation of bulk gravity in the domain wall solution~\eqref{dw} due
to the bulk and boundary potentials in the five--dimensional action.
That is, $\e$ measures the variation of the metric (and the dilaton)
as one moves across the orbifold.
Second, it measures the relative size $R/V$ of the orbifold and the
internal Calabi--Yau space. And third, it measures the relative
size of string--loop corrections to the four--dimensional
effective action which is indeed proportional to $T_R/S_R$. The
relation between loop corrections and bulk gravity is not
accidental. One can verify that the one loop corrections to the
four--dimensional effective action are in fact generated by the
non--trivial structure of the domain wall
solution~\cite{losw2}. Hence, the linear approximation in $\e$ implies
that we are considering a four--dimensional one--loop effective action
or, equivalently, five--dimensional bulk excitations that are
well approximated by linearized gravity. In the following, we will
use the term ``bulk excitations'' to mean this non--trivial orbifold
dependence induced by the potentials in the five--dimensional theory
and related to loop corrections. Note that, due to the $R^2$ corrections on
the boundaries mentioned above, higher--derivative corrections will
also induce a non--trivial orbifold dependence whenever those corrections
become relevant. It would be interesting to include those higher
derivative terms, specifically the boundary $R^2$ terms, in the
analysis. A related four--dimensional analysis with $R^2$ terms has
been performed in ref.~\cite{art}. Higher curvature terms in five
dimensions have been considered in ref.~\cite{ek}, however the
boundary $R^2$ terms were not included in the analysis of this paper.
As stated above, in this paper, we confine ourselves to the lowest
order in $\a '$.

Does the five--dimensional action~\eqref{S5} and, correspondingly, its
exact domain--wall solution~\eqref{dw} encode higher loop--effects
as well? Certainly it contains information beyond the one--loop level,
since the bulk potential in eq.~\eqref{S5} which is uniquely fixed
by five--dimensional supersymmetry is of the order $v^2\sim\e^2$.
However, higher--order corrections to the five--dimensional action
cannot be excluded. Therefore, while one expects higher loop
corrections to be described by bulk gravity effects and an action of
the type above, there might be modifications of the concrete
form~\eqref{S5} at higher order.
 
\section{Cosmological solutions}

Based on the above correspondence between the four-- and
five--dimensional theories we would now like to discuss the simplest
type of cosmological solutions, namely rolling radii solutions. These
solutions are characterized by an evolution of the universe driven by
kinetic energy and they provide superinflating as well as subluminally
expanding cosmological string backgrounds~\cite{pre_BB}. First, we would like
to review those solutions in the four--dimensional context.
Then we present new five--dimensional solutions that constitute the
generalization of rolling radii solutions to heterotic M--theory.
Furthermore, we discuss their relation to the known four--dimensional
solutions.

\vspace{0.2cm}

Let us first recall the conventional picture that arises in four dimensions.
We choose a four--dimensional metric of Friedmann--Robertson--Walker
type with flat spatial sections and scale factor $\a_4 = \a_4 (t_4)$, that is,
\begin{equation}
 ds_4^2 = g_{4\m\n}dx^\m dx^\n = -dt_4^2+e^{2\a_4}d{\bf x}^2\label{m4}\; .
\end{equation}
Accordingly, the other two fields are taken to be functions of time
only, that is, $\b_4 = \b_4 (t_4)$ and $\f_4 = \f_4 (t_4)$. Then the general
solution of the four--dimensional action~\eqref{S4} is of the form
\begin{equation}
 \a_4 = p_{4\a}\ln |t_4| + \bar{\a}_4\; ,\qquad \b_4 = p_{4\b}\ln |t_4|
  + \bar{\b}_4\; ,\qquad
 \f_4 = p_{4\f}\ln |t_4| + \bar{\f}_4\; ,\label{sf4}
\end{equation}
where $\bar{\a}_4$, $\bar{\f}_4$ and $\bar{\b}_4$ are arbitrary constants. The
expansion powers ${\bf p}_4 \equiv (p_{4\a} ,p_{4\b} ,p_{4\f} )$ are
subject to the two constraints
\begin{equation}
 3p_{4\a} - p_{4\f} = 1\; ,\qquad 9p_{4\b}^2+4p_{4\f} +2p_{4\f}^2=4\;
 .
 \label{cond4}
\end{equation}
Apart from trivial integration constants such as $\bar{\a}_4$,
$\bar{\b}_4$ and $\bar{\f}_4$, we therefore have a
one--parameter family of solutions specified by the solutions to
eq.~\eqref{cond4}. Generically, the scale factor of
the universe as well as both moduli fields evolve in time. As usual,
for each set of allowed expansion coefficients, we have a solution in
the negative--time branch, that is, for $t_4<0$ and a solution in
the positive--time branch, that
is for $t_4>0$. As stands, the former evolves into a future curvature
singularity while the latter arises from a past curvature singularity.
Frequently, for the discussion of superinflating cosmology, specific
solutions are chosen from the set specified by~\eqref{cond4}.
These specific solutions are characterized by a constant T--modulus
and, hence, by the expansion coefficients
\begin{equation}
 {\bf p}_4^{(T)} = \left(\pm\frac{1}{\sqrt{3}},0,\pm
                            \sqrt{3}-1\right)\; .
\end{equation}
As discussed in the previous section, the quantity $\e$, defined in
eq.~\eqref{epsilon}, is of particular importance in our context as it
measures the size of the four--dimensional loop corrections as well as
the five--dimensional gravitational bulk excitations. Going back to
the general class of solutions, we have from eq.~\eqref{epsilon}
and~\eqref{sf4} that
\begin{equation}
 \epsilon\sim |t_4|^{p_{4\b} -p_{4\f}}\; .\label{eps4}
\end{equation}
Generally, therefore, $\epsilon$ will be time--dependent. However, we
can ask if there are special solutions in the above set for which
$p_{4\b} = p_{4\f}$ and, hence, $\epsilon$ is constant. Such solutions
indeed exist and are characterized by the expansion powers
\begin{equation}
 {\bf p}_4^{(\e )} = \left(
   \frac{3}{11}\left( 1\pm\frac{4}{3\sqrt{3}}\right) ,
   \frac{2}{11}\left( -1\pm 2\sqrt{3}\right) ,
   \frac{2}{11}\left( -1\pm 2\sqrt{3}\right)\right)\; .
 \label{p4e}
\end{equation}
While, in the following, we will work with the general set of
solutions, we will comment on these special cases where appropriate.
After this review of the four--dimensional solutions, let
us now move on to the five--dimensional case.

\vspace{0.2cm}

Our goal is to specify the five--dimensional origin of the
above rolling radii solutions. That is, we would like to find the
solutions of the
five--dimensional theory~\eqref{S5} that, in the small--momentum
limit, reduce to the four--dimensional rolling radii solutions.
From the action~\eqref{S5}, it is clear that those solutions, in
addition to time, must depend on the orbifold coordinate $y$, as long
as the constant $v$ is non--zero. In fact, while models with $v=0$
exist~\cite{symm}, generically $v$ is non--vanishing. As a consequence,
exact cosmological solutions of the action~\eqref{S5} are not easy to
find. The first example has been given in ref.~\cite{cosm} using
separation of variables and we will come back to this example later
on. Some generalizations, also based on separation of variables,
including those with curved three--dimensional spatial section have
been presented subsequently~\cite{real}. Exact, non--separating
solutions have been found for a related action set up to describe the somewhat
different physical situation of potential--driven inflation within
M--theory~\cite{infl}. However, exact non--separating solutions for the
action~\eqref{S5} are hard to find and not a single example is known
to date. In this paper, we will, therefore, content ourselves with
giving approximate non--separable solutions. Such solutions can be
obtained by ``lifting up'' the four--dimensional rolling radii
solutions to five dimensions using the correspondence~\eqref{redans}
between the four-- and five--dimensional theories. Concretely, by
inserting~\eqref{m4} and \eqref{sf4} into eq.~\eqref{redans}, we obtain
as the approximate solution of the five--dimensional action~\eqref{S5}
\begin{equation}
 ds_5^2 = (1+\e h/3)\left(-dt_5^2+e^{2\a_5}d{\bf x}^2\right)
           +(1+4\e h/3)e^{2\b_5}dy^2\; , \qquad
 e^\f = e^{\f_5}(1 + \e h)\label{lin}
\end{equation}
where we have introduced the five--dimensional ``comoving'' time $t_5$
by
\begin{equation}
 dt_5^2 = e^{-\b_5-\f_5}dt_4^2\; .
\end{equation}
The five--dimensional scale factors $\a_5$, $\b_5$ and $\f_5$ show a
power--law behavior
\begin{equation}
 \a_5 = p_{5\a}\ln |t_5| + \bar{\a}_5\; ,\qquad \b_5 = p_{5\b}\ln |t_5|
  + \bar{\b}_5\; ,\qquad
 \f_5 = p_{5\f}\ln |t_5| + \bar{\f}_5\; ,\label{sf5}
\end{equation}
similar to the one in four dimensions. The expansion coefficients
${\bf p}_5=(p_{5\a},p_{5\b},p_{5\f})$ are subject to the constraints
\begin{equation}
 3p_{5\a}+p_{5\b} = 1\; ,\qquad 8p_{5\b}^2-4p_{5\b}+3p_{5\f}^2=4
 \label{c5}
\end{equation}
and can be obtained from their four--dimensional counterparts using
the relations
\bea
 p_{5\a} &=& \frac{2p_{4\a}-p_{4\b}-p_{4\f}}{2-p_{4\b}-p_{4\f}} \nn\\
 p_{5\b} &=& \frac{2p_{4\b}}{2-p_{4\b}-p_{4\f}} \label{coeff_rel}\\
 p_{5\f} &=& \frac{2p_{4\f}}{2-p_{4\b}-p_{4\f}}\; .\nn
\eea
We recall that the function $h=h(y)$ is defined by
\begin{equation}
 h(y) = \frac{|y|}{\p\r}-\frac{1}{2}\; .
\end{equation}
The all--important strong coupling expansion parameter $\e$, defined
in eq.~\eqref{epsilon}, is expressed in terms of five--dimensional
quantities as
\begin{equation}
 \e = \e_0e^{\b_5-\f_5}\sim |t_5|^{p_{5\b}-p_{5\f}}\; .\label{e5}
\end{equation}
We have now found new approximate
solutions of the five--dimensional theory that, via the
relations~\eqref{coeff_rel} between the expansion coefficients, are in
one--to--one correspondence with the four--dimensional rolling radii
solutions given in \eqref{m4}, \eqref{sf4}. Hence, as is the case for their
four--dimensional counterparts, these five--dimensional solutions
constitute a one--parameter set specified by the solutions to the
constraints~\eqref{c5}. While the lifting
procedure from four dimensions makes it rather easy to obtain those
solutions, they are quite non--trivial from a five--dimensional
viewpoint. In particular, they are generically non--separating, that
is, the time-- and orbifold--dependence do not generically factorize.
This can, for example, be seen from the function $e^{2\a_5}(1+\e h/3)$
that multiplies the three--dimensional spatial part of the
metric~\eqref{lin}. Here, the time--dependence resides in $\a_5$ and
$\e$ while the orbifold--dependence is encoded in $h$. Hence, as long
as $\e$ does depend on time (which it generically does), the variables do
not separate. From the discussion of the
previous section, the approximation that led us to those solutions is
valid as long as higher--derivative terms are negligible and, hence,
the momenta $\dot{\a_5}$, $\dot{\b_5}$ and $\dot{\f_5}$ have to be
sufficiently small. Furthermore, the expansion parameter $\e$ has to
be less than one. The above solutions are direct generalizations of the rolling
radii solutions to five dimensions. Apart from 
$\a_5$, $\b_5$ and $\f_5$ that describe the overall scaling of the
domain--wall configuration, there is also a less trivial dependence on
time through the expansion parameter $\e$. This dependence implies
that the size of transverse gravity excitations (the linear slope
in $y$) varies with time as well.

\vspace{0.2cm}

Can the above approximate five--dimensional solutions be promoted to
exact solutions of the action~\eqref{S5}? The simplest approach
to finding such exact solutions is clearly separation of variables. In
fact, in ref.~\cite{cosm} it was shown that the only separable
solutions (assuming a flat three--dimensional spatial universe) for
the action~\eqref{S5} are precisely of the form
\begin{equation}
 ds_5^2 = \left( 1+\frac{1}{3}\e h\right)\left( -dt_5^2+e^{2\a_5}
          d{\bf x}^2\right) +\left( 1+\frac{1}{3}\e h\right)^4
          e^{2\b_5}dy^2\; , \qquad
 e^\f =  \left( 1+\frac{1}{3}\e h\right)^3 e^{\f_5} \label{separable}
\end{equation}
where the scale factors $\a_5$, $\b_5$ and $\f_5$ evolve according to the
general power law~\eqref{sf5}. However, for the above to be an exact solution
the particular values
\begin{equation}
 {\bf p}_5^{(\e )} =
 \left(\frac{3}{11}\left(1\mp\frac{4}{3\sqrt{3}}\right),
 \frac{2}{11}(1\pm 2\sqrt{3}),\frac{2}{11}(1\pm 2\sqrt{3})\right)\; .
 \label{p5e}
\end{equation}
for the expansion coefficients must be chosen. These particular
coefficients satisfy the constraints~\eqref{c5}. Therefore, upon linearizing
the exact solutions~\eqref{separable} in $\e$ we recover particular
cases of our approximate solution~\eqref{lin}. This implies that, from
our one--parameter set of approximate solutions, exactly two can be
promoted to exact separating solutions while all other exact solutions
have to be non--separating. There is another way to characterize the
two separating solutions. It can be verified, using the
map~\eqref{coeff_rel} and \eqref{p4e}, \eqref{p5e}, that the
separating solutions correspond to those four--dimensional solutions
with constant strong--coupling expansion parameter.
This can also be directly seen in five dimension using eq.~\eqref{e5}
and the fact that $p_{5\b}=p_{5\f}$ for the coefficients~\eqref{p5e}.
Hence, we have found that the exact separable
solutions to our five--dimensional action are precisely those for which
the strong--coupling expansion parameter is constant in time, that is
\begin{equation}
 \e = \mbox{const}\; .
\end{equation}
Recalling the interpretation of $\e$ from the previous section, those
exact separable solutions can, therefore, be characterized as precisely
the ones for which the ratio of Calabi--Yau and orbifold
volumes is constant. Equivalently, they are precisely the ones for which
string loop corrections or excitations of bulk gravity are
constant in time. If, on the other hand, these quantities vary
in time, the corresponding exact solution is non--separable. For this
case, no exact explicit solution has been found yet and it may well be that
this can only be achieved using numerical methods.

\section{Role of the fifth dimension}

We would now like to discuss the results obtained so far, particularly
in view of a kinetic--energy driven phase of inflation and the role of
the fifth dimension in such a context. 

A solution of the usual problems of standard cosmology requires the
scale factor $a=e^{\a}$ of the three--dimensional space to
accelerate for some period in the early universe. Such a superluminal
evolution is realized precisely if~\cite{pre_BB,gv}
\begin{equation}
 \mbox{sign}(\ddot{a}) = \mbox{sign}(\dot{a})\; .\label{acc}
\end{equation}
where the dot denotes the derivative with respect to comoving time.
The condition~\eqref{acc} is frame--independent, as it should be. In
particular, it can be used either in the four--dimensional string
frame or the five--dimensional Einstein frame. Consequently, we have
omitted the subscripts specifying the frame. In general, one expects
inflating ($\dot{a}>0$) as well as deflating ($\dot{a}<0$) solutions
of eq.~\eqref{acc}, both of which are suited to solve the problems of
standard cosmology~\cite{gv}. In fact, the sign of $\dot{a}$, and hence the
notion of expansion and contraction, is not frame--independent. If
eq.~\eqref{acc} is not satisfied the evolution is decelerated or
subluminal. As before there are two cases, namely decelerated
expansion ($\dot{a}>0$) and decelerated contraction ($\dot{a}<0$).

\begin{figure}[ht]
   \centerline{\epsfig{figure=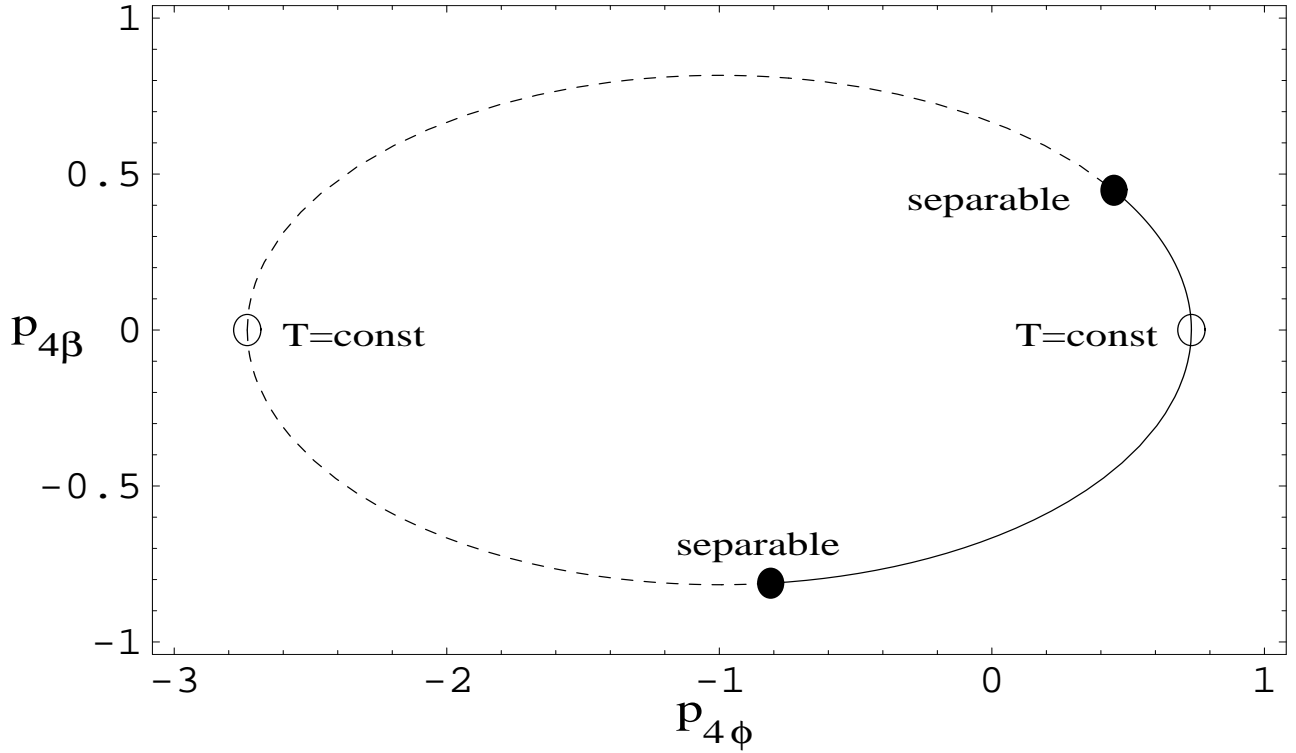,height=17cm,width=10cm,angle=-90}}
   \caption{The dilaton and T--modulus expansion coefficients in the
            four--dimensional string frame for the
            solutions given in the text. The solid dots correspond to the
            separable solutions, the circles to the solutions with
            constant T--modulus. The solid line represents solutions
            with $p_{4\b}-p_{4\f}<0$, the dashed line represents
            those with $p_{4\b}-p_{4\f}>0$.}
   \label{fig}
\end{figure}

For the solutions given in the previous section, it is easy to verify
that the condition~\eqref{acc} is satisfied as long as one chooses the
time to be negative. In other words, the complete one--parameter set
of solutions leads to accelerated evolution in the negative--time
branch. In the positive time branch, on the other hand, the
condition~\eqref{acc} is never satisfied. The evolution is, therefore,
always decelerated. Let us assume in the following
discussion that $t_4<0$. A convenient way to represent the set of
solutions is to plot their expansion powers.
This has been done in Fig.~\ref{fig} using the coefficients $p_{4\f}$
and $p_{4\b}$ for the dilaton and the T--modulus in the
four--dimensional string frame, subject to the second condition
in~\eqref{cond4}.
Let us now discuss the time evolution for the solutions represented in
Fig.~\ref{fig} in the negative--time branch. We start at
$t_4\rightarrow -\infty$, assuming an effective four--dimensional
description at this time. All solutions will, of course, eventually
develop large higher--derivative ($\a '$) corrections as
$t_4\rightarrow 0$. For example, the product of the ``momenta'' $\dot{\a_4}$,
$\dot{\b_4}$ and $\dot{\f_4}$ times the orbifold size is proportional
to $|t_4|^{p_{4\b}-1}$. This increases as $t_4\rightarrow 0$ since
$|p_{4\b}|<1$ always. The precise time when the lowest order $\a '$
approximation is invalidated depends, of course, on initial conditions. 

In section 2 we have discussed another sense in which the fifth
dimension may become relevant. Namely, the parameter $\e$ and, hence,
the excitation of fields in the fifth dimension may become large. At
the same time, this implies large loop corrections. As
we have seen, $\e\sim |t_4|^{p_{4\b}-p_{4\f}}$ and, therefore, its qualitative
behavior depends on the sign of $p_{4\b}-p_{4\f}$. Consequently, unlike the
higher--derivative ($\a '$) corrections discussed above, $\e$ does not
always increase in time. Instead, we should distinguish the three
cases (for the negative--time branch)
\begin{itemize}

\item $p_{4\b}-p_{4\f} > 0$~: Then $\e$ decreases in time, indicating
      decreasing bulk excitations/loop corrections. The Calabi--Yau
      space expands faster than the orbifold. Solutions with
      this property are represented by the dashed line in
      Fig.~\ref{fig}.

\item $p_{4\b}-p_{4\f} = 0$~: Then $\e = $ const, corresponding to
      constant bulk excitations/loop corrections. The Calabi--Yau
      space expands at the same rate as the orbifold. As discussed this
      case corresponds precisely to the two exact separable solutions
      that can be found. These solutions are indicated by the dots in
      Fig.~\ref{fig}. 

\item $p_{4\b}-p_{4\f} < 0$~: Then $\e$ increases in time indicating
      increasing bulk excitations/loop corrections. The orbifold
      expands faster than the Calabi--Yau space. The corresponding
      solutions are represented by the solid line in Fig.~\ref{fig}.

\end{itemize}

We see that bulk excitations in the fifth dimension are irrelevant
in the first two cases, even as we approach the singularity at
$t_4\rightarrow 0$. Of course, the system will still run into a large
curvature regime close to the singularity. We note that the
``standard'' solution with a constant T--modulus and inflation in the
$D=4$ string frame corresponds to the left circle in Fig.~\ref{fig}.
Hence, this solution falls into this category. The right circle, on
the other hand, corresponds to a deflating solution in the $D=4$
string frame and it falls into the third category. 

In general, in this third case, bulk excitations become
relevant close to the singularity. Whether that happens before or
after the systems enters the large curvature regime depends on
initial condition. Let us assume that we first enter a large $\e$
regime while higher derivative corrections are still small.
Then, while $\e$ grows, our approximate five--dimensional solution~\eqref{lin}
quickly becomes invalid. We know that the exact solutions that
govern the further evolution have to be non--separating.
Consequently, the time evolution and the excitation of bulk modes will
be entangled in a complicated way. As we have discussed, we expect
this to be described by a five--dimensional action of the
type~\eqref{S5} possibly with additional higher order corrections.
It would, therefore, be interesting to study exact non--separating
cosmological solutions of the action~\eqref{S5} in the region of large
$\e$. Unfortunately, analytic expressions for those solutions are
not available and numerical methods might be required.

However, we may try to extract some information about the behavior at
large $\e$ by looking at the four--dimensional metrics that, for a
given five--dimensional cosmological solution, are induced on the two
boundaries. For example, it would be of interest to know whether or
not a solution which is accelerated for small $\e$ can, when $\e$ is
large, smoothly become decelerated. The answer, unfortunately, is
negative, as we now demonstrate. Following ref.~\cite{infl}, let us write a
five--dimensional solution in the general form
\begin{equation}
 ds_5^2 = -e^{2\n}dt_5^2+e^{2\a}d{\bf x}^2+e^{2\b}dy^2\; ,\label{m5}
\end{equation}
were $\n$, $\a$ and $\b$ are functions of $t_5$ and $y$. Furthermore,
we take the dilaton $\f$ to be a function of $t_5$ and $y$. The
equations of motion for such an ansatz, following from the
action~\eqref{S5}, have been presented in
ref.~\cite{infl}. Particularly useful for the present purpose is the
$55$ component of the Einstein equation which reads explicitly
\begin{equation}
 3e^{-2\n}(\ddot{\a}-\dot{\n}\dot{\a}^2)-3e^{-2\b}({\a '}^2+\n '\a ')
  = -\frac{1}{4}e^{-2\n}\dot{\f}^2-\frac{1}{4}e^{-2\b}{\f '}^2+
    \frac{1}{6}v^2e^{-2\f}\; .\label{G55}
\end{equation}
Here the dot (prime) denotes the derivative with respect to $t_5$ ($y$).
Furthermore, working in the boundary picture, the functions in the
above ansatz have to satisfy the following conditions~\cite{infl}
\begin{equation}
 e^{\f-\b}\n '\left.\right|_{y=y_i}=e^{\f-\b}\a '\left.\right|_{y=y_i}
 = \frac{\sqrt{2}}{6}v\; ,\qquad e^{\f-\b}\n '\left.\right|_{y=y_i}=
   \sqrt{2}v\; ,\label{bcond}
\end{equation}
at the first (second) boundary at $y_1=0$ ($y_2=\p\r$).
These conditions arise as a consequence of the $Z_2$ orbifolding
and the boundary potentials in the five--dimensional
action~\eqref{S5}, as usual. Restricting eq.~\eqref{G55} to either one
of the boundaries, and using the conditions~\eqref{bcond}, it is easy
to show that
\begin{equation}
 \ddot{\a}_i-\dot{\n}_i\dot{\a}_i+2\dot{\a}_i^2=-\frac{1}{12}\dot{\f}_i^2\; .
 \label{fb}
\end{equation}
Here the subscript $i$ denotes the value of the respective field at
the boundary $i$, that is, for example $\a_i(t_5)=\a (t_5,y_i)$. We note that
the various potential terms occurring in the Einstein equation and the
boundary conditions~\eqref{bcond} cancel in this relation. As a
consequence, we have no unusual, linear relationship between the
Hubble parameter and the boundary stress energy in eq.~\eqref{bcond}.
The possibility of such unconventional relations has been first
observed in ref.~\cite{infl}. As a check, we can now verify that the
relation~\eqref{fb} is satisfied by our approximate five--dimensional
solutions. Putting eq.~\eqref{lin} in the form~\eqref{m5} and
restricting to the boundaries, we can read off the following expressions
\begin{equation}
 \a_i=p_{5\a}\ln |t_5|\mp\frac{1}{12}\e\; ,\qquad
 \f_i=p_{5\f}\ln |t_5|\mp\frac{1}{2}\e\; ,\qquad
 \n_i=\mp\frac{1}{12}\e
\end{equation}
where the upper (lower) sign refers to the boundary $i=1$ ($i=2$).
Here the expansion coefficient $p_{5\a}$, $p_{5\b}$ and $p_{5\f}$
satisfy the relations~\eqref{c5}. Inserting these expressions
and using~\eqref{c5}, we can indeed verify
that eq.~\eqref{fb} is satisfied to linear order in $\e$, as it should
be. We can now go further and use the relations~\eqref{fb} to deduce
properties of the solutions at arbitrary $\e$. In doing so we have to
be careful, of course, since presumably not every set of
fields $(\a_i,\n_i,\f_i)$ satisfying~\eqref{fb} can be
extended to a full five--dimensional solution. However, conversely, every
five--dimensional solution gives rise to induced fields on the
boundaries that do satisfy eq.~\eqref{fb}. It is this latter
connection that we are going to use. We introducing the boundary Hubble
parameters $H_i=\dot{\a}_i$ and choose the five--dimensional time
coordinate $t_5$ such that is becomes comoving time upon restriction
to the boundaries. This implies $\n_i=0$ and, hence, eq.~\eqref{fb}
can be written in the form
\begin{equation}
 \dot{H}_i=-\left(2H_i^2+\frac{1}{12}\dot{\f_i}^2\right)\; .\label{fb1}
\end{equation}
We conclude that $\dot{H}_i$ is always negative. Furthermore, the
criterion~\eqref{acc} for accelerated evolution can be brought into
the form $\mbox{sign}(\dot{H}_i+H_i^2)=\mbox{sign}(H_i)$. From
eq.~\eqref{fb1} we conclude that
$\dot{H}_i+H_i^2< 0$, always. Therefore, the evolution is accelerated
exactly if $H_i<0$. In this case, the boundaries deflate. On the other
hand, for expanding boundaries, $H_i>0$, the evolution must be
decelerated. Hence, a five--dimensional solution which changes
from acceleration to deceleration implies a transition from
$H_i<0$ to $H_i>0$ for the boundary Hubble rates. This, however,
cannot happen in a continuous manner since $\dot{H}_i<0$. We conclude
that a transition from acceleration to deceleration does not take
place, even for large values of $\e$. We note, however, that the
physically less interesting transition from deceleration
to acceleration is not excluded from the above argument. In
conclusion, we have shown that the solutions of our five--dimensional
theory do not evolve from acceleration to deceleration. This results
holds for arbitrarily large $\e$ corrections but only to lowest order
in $\a '$. It is quite conceivable that the inclusion of higher order
$\a '$ corrections can change this situation similarly to what happens
in four dimensions~\cite{art,ccm}. Some of those $\a '$ correction
arise on the boundaries of the five--dimensional theory and, hence,
lead to further bulk inhomogeneities. It would be interesting to
generalize the present work by including those corrections.

\vspace{0.2cm}

{\bf Acknowledgments}
A.~L.~is supported by the European Community
under contract No.~FMRXCT 960090. B.~A.~O.~is supported in part by 
DOE under contract No.~DE-AC02-76-ER-03071 and by a Senior 
Alexander von Humboldt Award.
 


\end{document}